\documentclass[a4paper,11pt]{article}
\usepackage[T1]{fontenc}
\usepackage[utf8]{inputenc}
\usepackage{lmodern}

\title{Is 1+1=2 an empirical proposition?}
\author{Yafet Sanchez Sanchez\footnote{E-mail:Y.SanchezSanchez@soton.ac.uk} \\ 
 Mathematical Sciences and STAG Research Centre, \\ University of Southampton,\\ Southampton,\\ SO17 1BJ } 
\date{\vspace{-5ex}}

\begin{document}
\maketitle

\section{Meaning in mathematical propositions.}

Imagine someone demands from us a proof that $1+1=2$.
Certainly, our reaction will depend on whether we are asked by a child or a
mathematician. While in the former case we can act by showing that one
coin put together with another coin, when counted again are two coins. In
the latter case, it is not clear what we are meant to show. Is a syntactic 
calculation in an abstract setting showing the truth about acausal, nontemporal,
and non-spatial objects what is required?

Consider again the child learning arithmetic. At what point, does she
grasp the same truth the mathematician is trying to prove in her theorems.
Or is it necessary to be familiar with Peano Axioms and first order logic
to understand the true meaning of $1+1=2$? We become bewitched by
what we believe lies behind the symbols. It seems like the meaning of the
mathematical proposition is covered by some fog and dispersing the fog is
the work of the mathematician.

But what is meaning in the first place? How do symbols, for example
words, acquire meaning? We follow Wittgenstein in ascribing meaning to a
large class of words as the way we use the word in language. For example,
the meaning of the word "table" is not a mental or physical representation
of a four legged object; the meaning of the word table is the way in which
we use it as in "Put the essay on the table" or "The data is shown in table
one". We can only say we understand the meaning of the sentence if, in the
former sentence we act by putting the essay on the table, and in the latter
by looking at the correct place. If someone says "I feel very table." we will
demand some explanation as the words are used in a manner which do not
fit any use that we are aware of. We do not understand the use of the word
table in that context. The meaning goes astray.

Can this idea of meaning be taken to mathematics? Or in other words: is
the use of the mathematical proposition what gives them meaning?
An analysis between empirical statements and mathematical propositions
is needed here. The truth of a mathematical proposition is independent
of any empirical phenomena. The use of mathematical prepositions supports
this premise. For example, if we see two drops of water coming together and
forming one, we are not tempted to claim that $1+1=2$ is false. Rather,
we conclude that this is not an example where we can use the proposition
$1+1=2$. Imagine that actually all objects behave in this peculiar
way by sometimes merging sometimes splitting such that the proposition
$1+1=2$ can never be used in the physical world. In that case, the
truth or falseness of arithmetic propositions will not matter anymore as the
whole use of arithmetic will reduce to that of a game of symbols. While the
truth or falseness of mathematical proposition are independent of empirical
phenomena, the sense (or meaning) of mathematical propositions is not.

The fact is that nature does not behave in this way and that empirical
regularities appear constantly. We as humans become aware of this brute
fact and act accordingly. We employ initially the symbols to show the regularity
and then "harden" them into mathematical propositions. This is the
mathematical practice. The mathematicians job is to see that a stick used to
measure the length of the objects can become a ruler. The usefulness then
of mathematics relies on us behaving the same way when presented with
the same ‘mathematically’ related situations (arranging, sorting, recognising
shapes, performing one-to-one correspondences, and so forth.) This is
not an agreement of opinion, but an agreement in what being a human is.
This is what gives mathematics its objectivity.

\section{Physics is an empirical science.}

Physics is the empirical science concerned with the behaviour of inanimate
matter. The term empirical science refers to the fact that physics relies
solely in the ability to test claims about the world with an experiment.
Nevertheless, physics is a word and therefore has different meanings depending the use of it in different contexts. For example, we can also describe
physics to be what physicist do. Before getting into a loop and describe a
physicist as someone who does physics, we can define a physicist as those
members of society which work in physics departments and receive funding
to do physics. In this context, physics is defined as anything this group of
people do. These two meanings are interwoven but they are not equivalent.
Our first definition describes an activity, the second one describes an activity
done by certain people.

If we are interested in our first definition of physics, we may ask: how is
theoretical physics possible with the picture of mathematical propositions
presented above? A problem seems to appear: if mathematics are
hardened empirical regularities what are the mathematical symbols theoretical physicist use
to describe ‘yet to be found’ empirical regularities?

The problem can be resolved by changing our view toward the tools a theoretical physicist use. The objective of a theoretical physicist is not only to find rules that agree with empirical data
but ways to find new empirical data. The way one achieves this is nowhere regulated. Therefore, theoretical physics is a no-man’s-land in the junction
of mathematics, physics and philosophy where everything is allowed as far
as one is able to make accurate predictions. In this sense the mathematical symbols a theoretical
physicist uses is an extremely useful and unregulated symbol game that differs from mathematics even if
the symbols used are the same.

If the more sociological point of view towards physics is taken then
theoretical physics is defined by the practitioners. Is this then just a matter
of agreement between a community? Compare with the mathematicians
accepting something is proven: what the community is agreeing to is not
an opinion but an agreement in the same way of acting (analogous to the way: we breath
the same, our hearts beat the same). Nevertheless, are the agreements
in the case of theoretical physicist equally grounded as in the case of the
mathematicians? The distinction between them is a philosophical task that
must be done in order to avoid misunderstanding. Mathematical meaning
can go astray too.

\section{Ordinal arithmetic and String theory.}
The discussion above provides initial ideas about how theoretical physics and
mathematics can be discussed in terms of use and meaning. The main shift
one would like to achieve is to move from the dichotomy of "true" and "false"
propositions to the notions of "sense" and "nonsense". What one needs to see
is the background (or context) which allow both activities to have meaning. In particular,
there are two examples which provide a formidable starting point of analysis:
"Ordinal Arithmetic" and "String theory".

Ordinal arithmetic seems to put heavy tension on the claims made before.
Where is the empirical regularity that makes ordinal arithmetic true? The
temptation to make this question must be avoided. We need to ask: "Where
is the empirical regularity that makes ordinal arithmetic to have sense?", or
"How is ordinal arithmetic used?". The analysis then should go to understand
the connections between ordinal arithmetic and the rest of mathematics.
Notice that the relevance of the concept of infinity for the discussion
has been shifted from a metaphysical discussion to a pragmatic one. The question we are interested in should be: How are we using infinity in order to
have useful statements in ordinal arithmetic? But, what if applications are
lacking and theorems do not have implications to any other part of mathematics.
The answer must be similar to: "What is the use of a cog that spins
vigorously but is not attached to the rest of the machine?"

String theory provides an excellent case of analysis because of the multiplicity
of meanings. It is hard to draw a line between what is mathematics
and what is theoretical physics in String theory. However the must important
part is to be sure that there is not mathematical symbols out of
the appropriate context and therefore nonsense. This is what the analysis
should do. The claims that string theory is a part of theoretical physics and
therefore an approach to describe and predict empirical phenomena
must be independent of the mathematical machinery developed. Arguing
otherwise is close to misunderstand the difference between physics and
mathematics (the difference as a human activity). In this case, string theory
as a theoretical physical tool can only be evaluated in the light of usefulness
to explain and predict empirical phenomena. An experiment is needed to
settle the question.

The claim that String theory contains mathematical propositions has
also to be carefully analysed. This means that any mathematical propositions in String
theory must be approached with the same mathematical attitude as when
one is doing mathematics. One can not use the freedom of the theoretical physicists to manipulate the symbols and then claim one has a rigorous mathematical proposition. That would lead to misunderstanding. However, there might be also the case that String
theory is a new kind of activity which is neither physics, nor mathematics.
Only the dialogue between philosophers, mathematicians and physicists can
shed light on this delicate matter.\\\\
\emph{
The work here presented has taken as a basis the ideas of Ludwig Wittgenstein
in \cite{phil} and \cite{remarks}.\\
Also the work in \cite{so} was influential for the work.}

\end{document}